# MPC-Based Operation Strategy for Electric Vehicle Aggregators Considering Regulation Markets

Liling Gong, Ye Guo, *Senior Member, IEEE*, and Hongbin Sun, *Fellow, IEEE*

*Abstract*—The optimal operation problem of electric vehicle aggregator (EVA) is considered. An EVA can participate in energy and regulation markets with its current and upcoming EVs, thus reducing its total cost of purchasing energy to fulfill EVs' charging requirements. A model predictive control (MPC) based optimization is developed to consider the future arrival of EVs as well as energy and regulation prices. The index of conditional value-at-risk (CVaR) is used to model the risk-averseness of an EVA. Simulations on a 2000-EV test system validate the effectiveness of our work in achieving a lucrative revenue while satisfying the charging requests from EV owners.

## Nomenclature

*Indices/Sets*

| | |
|---|---|
| $s/\Omega$ | Index/set of all the scenarios. |
| $V_1/V_2$ | Set of all V1G/V2G EVs during the operating day. |
| $\Xi_1/\Xi_2$ | Set of connected EVs in the V1G/V2G mode. |
| $\Lambda_1/\Lambda_2$ | Set of upcoming EVs in the V1G/V2G mode. |

*Parameters*

| | |
|---|---|
| $\lambda_\tau$ | Energy clearing price at time $\tau$. |
| $\mu_\tau$ | Regulation clearing price at time $\tau$: $\mu_\tau = \mu_\tau^{rc} + \mu_\tau^{rp} m_\tau^{RegD}$, where $\mu_\tau^{rc}/\mu_\tau^{rp}$ denote regulation capacity/performance price; $m_\tau^{RegD}$ denotes the mileage of AGC signals at $\tau$. |
| $\lambda_\tau^{(s)}/\mu_\tau^{(s)}$ | Energy/Regulation price at time $\tau$ in scenario $s$. |
| $\psi$ | Degradation price due to discharging. |
| $\pi_s$ | Occurrence probability of scenario $s$. |
| $H$ | Window width of the rolling-horizon optimization. |
| $K$ | Current operating time slot. |
| $\varphi/\varphi'$ | Penalty cost of a unit unfulfilled regulation capacity at time $K/K+1$, respectively. |
| $SoC_j^A$ | Arrival SoC of EV $j$. |
| $SoC_j^R$ | Required SoC of EV $j$. |
| $SoC_j^{max}$ | Maximum threshold SoC of EV $j$. |
| $SoC_j^{min}$ | Minimum threshold SoC of EV $j$. |
| $t_j^a/t_j^d$ | Arrival/Departure time of EV $j$. |
| $B_j$ | Rated energy capacity of EV $j$. |
| $p_j^{max}$ | Maximum power rating of EV $j$. |
| $\mathbf{A}_\tau^{\Xi/\Lambda}$ | Binary availability vector for an EV set at $\tau$: $\mathbf{A}_t^{\Xi_2,[j]}=1$, means the $j$-th EV in set $\Xi_2$ is unconnected at time $\tau$. |
| $\mathbf{\Gamma}_K^{\Xi/\Lambda}$ | Vector of remained parking time for a set at time $K$. |
| $\mathbf{P}_+^{\Xi_1/\Xi_2}$ | Vector of maximal power rating for the set $\Xi_1/\Xi_2$. |
| $\mathbf{P}_+^{\Lambda_1/\Lambda_2,(s)}$ | Vector of predicted maximal power rating for a set in $s$. |
| $\mathbf{E}_r^{\Xi_1/\Xi_2}$ | Vector of remained required energy for the set $\Xi_1/\Xi_2$. |
| $\mathbf{E}_r^{\Lambda_1/\Lambda_2,(s)}$ | Vector of predicted required energy for a set in $s$. |
| $\mathbf{E}_{+/-}^{\Xi_2}$ | Vector of upper/lower energy for set $\Xi_2$. |
| $\mathbf{E}_{+/-}^{\Lambda_2,(s)}$ | Vector of upper/lower energy for set $\Lambda_2$ in scenario $s$. |
| $R_K$ | Cleared regulation capacity bid of time $K$. |

*Variables*

| | |
|---|---|
| $x_{j,\tau}/y_{j,\tau}$ | Charging/Discharging POP of EV $j$ at time $\tau$. |
| $z_{j,\tau}$ | Regulation capacity of EV $j$ at time $\tau$. |
| $\mathbf{X}_K^{\Xi_1/\Xi_2}$ | Vector of charging POPs for the set $\Xi_1/\Xi_2$ at time $K$. |
| $\mathbf{Y}_K^{\Xi_2}$ | Vector of discharging POPs for the set $\Xi_2$ at time $K$. |
| $\mathbf{Z}_K^{\Xi_1/\Xi_2}$ | Vector of regulation capacity for the set $\Xi_1/\Xi_2$ at $K$. |
| $\mathbf{X}_\tau^{\Xi/\Lambda,(s)}$ | Vector of charging POPs for a set at time $\tau$ in $s$. |
| $\mathbf{Y}_\tau^{\Xi_2/\Lambda_2,(s)}$ | Vector of discharging POPs for a set at $\tau$ in $s$. |
| $\mathbf{Z}_\tau^{\Xi/\Lambda,(s)}$ | Vector of regulation capacity for a set at $\tau$ in $s$. |
| $R_{K+1}$ | Regulation capacity bid of time $K+1$ submitted by EVA. |
| $\omega_K/\omega_{K+1}^{(s)}$ | Amount of unfulfilled regulation capacity at time $K$ or at time $K+1$ in scenario $s$. |

## I. Introduction

The adoption rate of electric vehicles (EVs) has been accelerated during the last decade [1], and there still exists a strong momentum in the EV growth since they are proven to be an important component in our transition to a carbon-neutral society. It can not only reduce carbon emissions in the transportation sector, but also provide the power grid with crucial flexibility support [2]-[4], such as regulation services.

With rapid integrations of renewable energy sources, the power grid is facing more and more uncertainties and volatilities. At the same time, the portion of traditional sources of regulation services, thermal power plants, is dropping in order to ease the carbon emission. Therefore, the power grid is in desperate need of new types regulation resources. EVs on average spend 90% of the time in parking lots[5], being similar to grid-connected batteries. Thus there is a huge potential for

This paragraph of the first footnote will contain the date on which you submitted your paper for review, which is populated by IEEE. It is IEEE style to display support information, including sponsor and financial support acknowledgment, here and not in an acknowledgment section at the end of the article. For example, "This work was supported in part by the U.S. Department of Commerce under Grant BS123456."

Second B. Author, Jr., was with Rice University, Houston, TX 77005 USA. He is now with the Department of Physics, Colorado State University, Fort Collins, CO 80523 USA (e-mail: author@lamar.colostate.edu).

Third C. Author is with the Electrical Engineering Department, University of Colorado, Boulder, CO 80309 USA, on leave from the National Research Institute for Metals, Tsukuba 305-0047, Japan (e-mail: author@nrim.go.jp).

EVs to provide regulation supports to the grid.

However, it is impractical for an individual EV to participate in the wholesale market directly, because i) most electricity markets are carried out on an MW basis (the minimum regulation capacity offer in PJM is 0.1MW [6]); ii) it is difficult for ISO to manage such a large number of transactions in the system level. Thus, a new entity, electric vehicle aggregator (EVA), has been introduced to coordinate interactions between EV owners and ISO. Compared with battery storage systems, a key challenge to EVAs lies in the uncertain parking time of future EVs while accounting for the physical aspects of each individual EV. Moreover, satisfying charging energy requests of arrived EVs should be a prior concern of EVAs. A well-designed EVA operation scheme can optimize its economic benefits in electricity markets while satisfying the charging demands of all EVs in the aggregator.

In the coordination between EVs and ISO, the main tasks of an EVA include bidding in electricity markets and dispatching charging power of EVs to track issued regulation signals. To obtain a profit-maximizing regulation bid ahead-of-time, an effective aggregated EV model is highly desirable to evaluate the available capacities. In such a model, the individual EV charging behavior forecasting is not required. Reference [7] utilizes a state-space method to simplify the EVA model in which chargers in ON and OFF status. A generic and scalable aggregation approach for flexible energy systems based on zonotopic sets and Minkowski sum is introduced in [8], however, the time-dependent EV charging demand makes this method computationally expensive. In [9], an aggregate model is proposed for EV fleet which employs aggregated parameters to represent energy and power boundaries. This model performs well for the EV charging problem [10], however, it tends to overestimate the regulation capacity in the symmetrical regulation markets[11], like PJM. Therefore, it remains an open question how to construct an aggregated model to describe the EV flexibility when participating in regulation markets.

A real-time (RT) dispatching strategy is key to EVA operation, especially considering regulation provisions. The aggregator manages charging power of parked EVs to quickly respond to regulation signals sent by ISO (e.g., every 2 seconds in PJM). In [12], a RT greedy-index dispatching policy is proposed for the EVA to provide regulation by sending out charging, idling, or discharging commands to each EV, assuming the rated power to be used for both charging and discharging status. In [13], a hierarchical V2G control strategy is modeled to distribute the ACE-based regulation task among all EVs in charging stations. However, these strategies do not take into account the economic benefits of market participation. From an EVA perspective, the market-side profitability is an important diver for providing regulation, and therefore it should not be omitted when dealing with EVA issues.

Some dispatching strategies are incorporated into the EVA bidding problem to maximize its market profits. In [14], an RT algorithm based on linear programs with charging priority weights is developed to allocate the EVA regulation task to individual EVs. In [9], a Laxity-SoC-based smart charging criteria is adopted during EVA operation. However, since EV charging is a dynamic process that can be represented as a multi-time horizon problem, the RT commands made by these heuristic methods may have negatively impact on subsequent regulation capabilities. A challenge arises in EVA operations: how to pursue the optimum market profits while simultaneously achieving the delivery of regulation services and the charging requests of parked EVs?

To deal with the coupling effect of RT dispatching and market profit, model predictive control (MPC) method is presented to optimize multi-time decisions under additional constraints. In [15], a design of new V2H-HEMS aggregator that enables each household to participate in a regulation market is presented where MPC method is adopted to consider the individual consumption behavior and operating constraints. Reference [16] develops an RT charging controller, and MPC method is used to track AGC signals for which forecasts are available. An MPC scheme is proposed for EVAs to participate in regulation market in [17], where a prediction method based on a SARIMA model on the regulation prices is also implemented, but energy cost is ignored in that problem. However, these aforementioned methods do not take into account charging demands of upcoming EVs when optimizing their strategies, whereas MPC method can be valuable to describe such dynamics.

In this paper, we develop a rolling-horizon operation strategy for EVAs to participate in the RT energy and regulation markets. The EVA takes minimizing total cost as an objective, while considers both satisfying EV charging requests and fulfilling its cleared regulation bids. With the problem description in Section II, we formulate the determinist EVA operation problem in Section III, where EVA is modeled to be static, i.e., EVA perfectly knows the charging behavior of each EV in advance. We first study the optimal capacity offering for the EVA in both energy and regulation markets, assuming the dynamic regulation signal is energy neutral on average. By rigorous analysis, the optimal preferred operation point (POP) of an EV is always one of a few cases in order to reserve its capacity for regulation, as shown in lemma 1 and 2. Furthermore, a novel concept of charging flexibility index is defined. Accordingly, a large number of EVs with the same parameters of arrival/departure time and charging flexibility index can be equated to a virtual EV. As a result, a group of virtual EVs can be utilized to predict the aggregated capacity of EV fleet. In Section IV, we take into account the EVA dynamic with EVs coming and departure. Different time-scale controls from bidding to RT dispatching are proposed in the MPC scheme, which is necessary to optimize market benefits of the EVA while ensuring the charging demands of EV owners. A two-stage stochastic programming model is developed to handle the comprehensive uncertainties of energy and regulation prices as well as the future arrival of EVs. Meanwhile, conditional value-at-risk (CVaR) is leveraged in the model to make a risk-averse decision against the unforeseen prediction errors in the above parameters. We demonstrate the performance of the proposed scheme in Section V and conclude in Section VI.

## II. PROBLEM DESCRIPTION

The RT operation problem of an EVA is considered, where the EVA can participate in energy and regulation markets by managing a fleet of connected EVs, as illustrated in Fig. 1.

Without loss of generality, the market rules of this work are based on that of PJM. In PJM, all regulation offers must be submitted before 14:15 in the day-ahead stage; however, to accurately reflect each resource's availability during the operating day, the submitted offers can be adjusted on an hourly basis up until 65 minutes prior to the start of the operating hour [6, p.66]. Upward and downward regulation are treated as the same form of products; thus, a resource cleared in the regulation market must be able to adjust in both directions. There are two types of regulation signals in PJM: the traditional regulation signal (RegA) for resources with low ramping capability, and the dynamic regulation signal (RegD) for resources with high ramping capability but limited energy availability. In this work, the dynamic regulation signals is tracked by EVA due to the fast-ramping rate of in-vehicle batteries.

Our main assumptions in this paper are listed as follows:

1) Once an EV arrives, it will declare the departure time and expected SoC to the aggregator; and the EVA needs to fulfill the charging request of each individual EV before its leaving.

2) The length of a time slot is set as one hour, and all market transactions are conducted on an hourly basis. The EVA is allowed to change its offered regulation capacity one time slot before the operating hour, and it submits a quantity-only regulation capacity offer. It is assumed that the regulation capacity offered by EVA is always accepted by ISO at market clearing prices, and all energy consumed by EVA is fully cleared in the energy market.

3) Hourly energy-neutral dynamic regulation signal is assumed in our optimization. This is because the hourly-average RegD signal is a zero-mean signal with small variance [18], and the variance due to system-level bias is difficult to predict accurately for EVA. Moreover, RegD signals in PJM were originally designed to be energy neutral within 15 minutes [19].

4) Battery degradation cost due to discharging POP is considered, whereas that caused by regulation is ignored. Because the addition of regulation service on top of energy arbitrage should not substantially increase the capacity degradation rate of a battery operating in energy arbitrage alone [20].

## III. THE DETERMINISTIC EVA OPERATION PROBLEM

In this section, we present an ideal case, where all the non-causal information, including market clearing prices and stochastic EV charging behaviors, are known to the EVA. Two types of EV charging modes are considered: (i) V1G mode, where EVs charge power from the grid only. However, it can still offer regulation service by adjusting its charging power away from the POPs; (ii) V2G mode, where EVs are allowed to discharge power to the grid. Let $V_1$/$V_2$ denote the set of plug-in EVs in V1G/V2G mode in the aggregator during the operating day. In prior to introducing the EVA optimization problem, we define three important parameters, i.e., required energy of EV battery before departure, and maximum/ minimum energy of

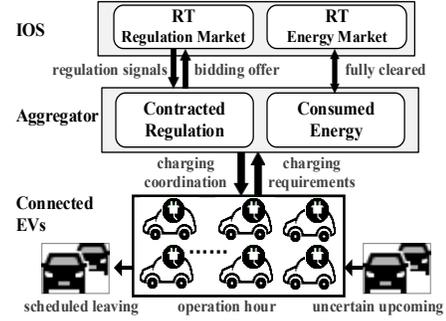

Fig. 1. Illustrative schematic of the EVA coordination between two sides.

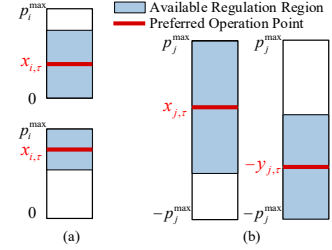

Fig. 2. Available frequency regulation capacity of individual EV at time τ. (a) V1G mode. (b) V2G mode

EV battery before reaching its safety bounds, which can be calculated, respectively, as follows:

$$E_i^r = \left(SoC_i^R - SoC_i^A\right)B_i, \quad \forall i \in V_1 \cup V_2. \quad (1)$$

$$E_i^{\max} = \left(SoC_i^{\max} - SoC_i^A\right)B_i - \rho p_i^{\max}, \quad \forall i \in V_2, \quad (2)$$

$$E_i^{\min} = \left(SoC_i^{\min} - SoC_i^A\right)B_i + \rho p_i^{\max}, \quad \forall i \in V_2, \quad (3)$$

where $B_i$, $p_i^{\max}$ denote the rated energy capacity and maximum power rating of EV $i$ battery; $\rho$ denotes the minimum fraction of energy reserved as a buffer for a unit regulation capacity. In (2) and (3), we reserve sufficient energy margin for each V2G EV to provide regulation services.

### A. EVA Operation Problem

1) V1G EV Constraints

As illustrated in Fig.2, available regulation region represents a symmetric band around the scheduled POP. For the $i$-th V1G EV, its charging POP $x_{i,\tau}$ and regulation capacity $z_{i,\tau}$ are constrained through:

$$x_{i,\tau} + z_{i,\tau} \leq p_i^{\max}, \quad \forall i \in V_1 \text{ and } \tau, \quad (4)$$

$$0 \leq z_{i,\tau} \leq x_{i,\tau}, \quad \forall i \in V_1 \text{ and } \tau. \quad (5)$$

With assumption 3, the scheduled energy level will not be affected too much by regulation over time. To ensure its charging request, charging POPs of EV $i$ in the parking time should satisfy:

$$\sum_{\tau=t_i^a}^{t_i^d} x_{i,\tau} = E_i^r, \quad \forall i \in V_1. \quad (6)$$

2) V2G EV Constraints

To incorporate battery degradation compensation, we model charging/discharging POP of EV $j$ seperatedly, denoted by $x_{j,\tau}$, $y_{j,\tau}$. The scheduled POP and committed regulation capacity cannot exceed the maximum charging/discharging rate $p_j^{\max}$:

$$x_{j,\tau} + z_{j,\tau} \leq p_j^{\max}, \quad \forall j \in V_2 \text{ and } \tau, \quad (7)$$

$$y_{j,\tau} + z_{j,\tau} \leq p_j^{\max}, \quad \forall j \in V_2 \text{ and } \tau, \quad (8)$$

$$x_{j,\tau}, y_{j,\tau}, z_{j,\tau} \geq 0, \quad \forall j \in V_2 \text{ and } \tau. \quad (9)$$

Similarly, the required energy of EV $j$ before its departure is given in (10); in (11), we guarantee that the energy level of EV $j$ remains in its permissible range at any $\tau$:

$$\sum_{\tau=t_j^a}^{t_j^d} (x_{j,\tau} - y_{j,\tau}) = E_j^r, \quad \forall j \in V_2. \quad (10)$$

$$E_j^{\min} \leq \sum_\tau (x_{j,\tau} - y_{j,\tau}) \leq E_j^{\max}, \quad \forall j \in V_2 \text{ and } \tau, \quad (11)$$

*3) EVA Objective Function*

The total cost considering the revenues from regulation participation, the energy cost and the battery degradation cost is minimized over the EVA operating day:

$$\min_{\substack{x_{i,\tau}, z_{i,\tau} \\ x_{j,\tau}, y_{j,\tau}, z_{j,\tau}}} \left[ \sum_{i \in V_1} \left( \sum_{t_i^a \leq \tau \leq t_i^d} \lambda_\tau x_{i,\tau} - \mu_\tau z_{i,\tau} \right) + \sum_{j \in V_2} \left( \sum_{t_j^a \leq \tau \leq t_j^d} \lambda_\tau (x_{j,\tau} - y_{j,\tau}) - \mu_{j,\tau} z_{j,\tau} + \psi y_{j,\tau} \right) \right]. \quad (12)$$

where $\psi$ denotes the battery degradation price caused by its discharging: $\psi>0$ always holds, thus, the optimal charging and discharging POPs at any $\tau$ cannot be positive simultaneously.

*B. Optimality Analysis*

After solving this deterministic problem, optimal scheduling of all EVs can be obtained. In particular, optimal solutions to a V1G EV $i$ can be denoted by $[(x_{i,\tau}^*, z_{i,\tau}^*), \forall \tau]$; optimal solutions to a V2G EV $j$ can be denoted by $[(x_{j,\tau}^*, y_{j,\tau}^*, z_{j,\tau}^*), \forall \tau]$. For an EVA, its optimal capacity offerings during the operating day can be calculated through one time optimization, as follows:

$$E_\tau = \left[ \sum_{i \in V_1} x_{i,\tau}^* + \sum_{j \in V_2} (x_{j,\tau}^* - y_{j,\tau}^*) \right], \quad \forall \tau, \quad (13)$$

$$R_\tau = \left[ \sum_{i \in V_1} z_{i,\tau}^* + \sum_{j \in V_2} z_{j,\tau}^* \right], \quad \forall \tau. \quad (14)$$

In (13), the total energy consumption of EVA is modeled: if this value is negative at time $\tau$, then the EVA will sell energy back to the grid. The optimal regulation capacity bid is presented in (14), which should be submitted to ISO by the EVA.

*1) Properties of EVs in V1G Mode*

*Lemma 1*: For any V1G EV $i \in V_1$, if the minimization problem has a solution, there must exist an optimal solution that satisfies:

$\forall i \in V_1$ and $\tau$,

$$x_{i,\tau}^* = \left\{ 0, \ p_i^{max}, \ \frac{p_i^{max}}{2}, \ E_i^r - \left\lfloor \frac{2 \times E_i^r}{p_i^{max}} \right\rfloor \text{(when } \tau = \chi) \right\},$$

$$z_{i,\tau}^* = \min\left\{ x_{i,\tau}^*, \ p_i^{max} - x_{i,\tau}^* \right\}.$$

*Proof*: See Appendix A. ∎

In general, the optimal charging POP of an EV in the V1G mode is always one of the three cases: no charging, full charging and half charging, except at most one specific time interval to satisfy the residual energy demand ($E_i^r - \lfloor 2 \times E_i^r / p_i^{max} \rfloor$). And we refer to this special time slot as the marginal time slot $\chi$.

*Definition 1*: The charging flexibility index of a V1G EV $i$ is defined to be an integer as follows:

$$F_i^{V1G} = \left\lfloor \frac{2 \times E_i^r}{p_i^{max}} \right\rfloor. \quad (15)$$

*Theorem 1*: If a set of V1G EVs (where $i=1,\cdots,N$) have the same $t_i^a, t_i^d, F_i^{V1G}$ values, the optimal aggregated solutions $\{\Sigma_i x_{i,\tau}^*, \Sigma_i z_{i,\tau}^*: \forall \tau\}$ can be equivalent to that of a virtual V1G EV who owns the same $t_i^a, t_i^d$. Other parameters of this virtual EV are:

$$E_v^r = \sum_{i=1}^N E_i^r, \ p_v^{\max} = \sum_{i=1}^N p_i^{\max}. \quad (16)$$

*Proof*: See Appendix B. ∎

According to theorem 1, we can equate any number of EVs with the same characteristics to one EV, greatly improving the computational efficiency.

*Remark 1*: Let $\sigma$ denote the regulation compensation price of for a unit regulation capacity, which EVA pays to EV owners for the market profit share. Then, objective function of EV $i$ in (12) changes to $\{\min \sum_\tau [\lambda_\tau x_{i,\tau} - (\mu_\tau - \sigma) z_{i,\tau}]\}$. The EVA has the same $\sigma$ for all EVs. If $\mu_\tau \geq \sigma, \forall \tau$ holds, theorem 1 is still valid.

*2) Properties of EVs in V2G Mode*

Although the replacement cost of in-vehicle batteries is trending downward, it is still expensive. The current cost for commercially-available lithium-ion batteries is between $150 and $300 per kWh. If we take the battery cost of $200/kWh as an example, and there are 2000 life cycles in 80% DoD, then the battery degradation price is $200/(2000*0.8) per kWh = 125 [$/MWh]. Settlements verified hourly energy prices and regulation prices in PJM for the last quarter of year 2021, is shown in Fig.3. And the average energy/regulation prices in 2021 are 38.2 and 30.1 [$/MWh], respectively. Here, we add another assumption of prices which seems to hold true in current status.

*Assumption 5*: Throughout that all market clearing prices of energy and regulation satisfy the situation $\{\mu_\tau + \psi > \lambda_\tau: \forall \tau\}$, where the regulation market price $\mu_\tau$ plus battery degradation price $\psi$ can be always larger than the energy market price $\lambda_\tau$.

*Lemma 2*: For any V2G EV $j \in V_2$, if the minimization problem has a solution, there must exists an optimal solution that satisfies:

$\forall j \in V_2$ and $\tau$,

$$x_{j,\tau}^* = \left\{ 0, \ p_j^{max}, \ E_j^r - \left\lfloor \frac{E_j^r}{p_j^{max}} \right\rfloor \text{(when } \tau = \chi) \right\},$$

$$y_{j,\tau}^* = 0, \quad z_{j,\tau}^* = \min\left\{ x_{j,\tau}^*, \ p_j^{max} - x_{j,\tau}^* \right\}.$$

*Proof*: See Appendix C. ∎

As it can be observed in Fig.3, there is a strong correlation between the energy price and regulation price fluctuations. V2G vehicles try to optimize the total revenue with full consideration of the differences between energy and regulation prices. And under our price assumption, the optimal charging power for a V2G EV is always one of the two cases: no charging, and full charging, except at most one marginal time interval.

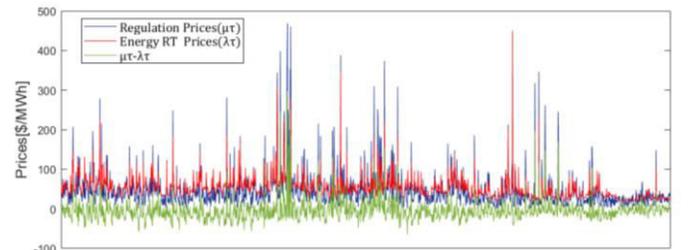

Fig. 3. Historical hourly prices for the last quarter of year 2021 in the PJM.

*Definition* 2: The charging flexibility index of a V2G EV $j$ is defined to be an integer as follows:

$$F_j^{V2G} = \left\lceil \frac{E_j^r}{p_j^{\max}} \right\rceil. \quad (17)$$

*Theorem* 2: If a set of V2G EVs (where $j=1,\cdots,M$) have the same $t_j^a$, $t_j^d$, $F_j^{V2G}$ values, the optimal aggregated solutions $\{\Sigma_j x_{j,\tau}^*, \Sigma_j y_{j,\tau}^*, \Sigma_j z_{j,\tau}^*: \forall \tau\}$ are equivalent to that of a virtual V2G EV who owns the same $t_j^a$, $t_j^d$. Other needed parameters of this EV are:

$$E_v^r = \sum_{j=1}^{M} E_j^r, \; p_v^{\max} = \sum_{j=1}^{M} p_j^{\max}. \quad (18)$$

*Proof*: See Appendix D. ∎

*Remark 2:* With above theorems, we can aggregate EVs into a virtual EV set with respect to arrival/departure times and charging flexibility index values, which will reduce the computational time and complexity. Note that slots in system time is finite and likewise the integer charging flexibility index is also finite due to the nature of EV batteries. Thus, the number of EVs in this virtual set will be limited regardless of how many EVs parking physically in the EVA.

## IV. THE MPC-BASED EVA OPERATION SCHEME

Discussions in the last section assume perfect predictions. However, an important challenge to the EVA operation lies in the fact that the system is dynamic with EVs coming and departing all the time. Moreover, it is a multi-scale problem with the sequentially revealed uncertainties in both market prices and upcoming EVs, where the EVA must satisfy all charging requests from connected EVs at their departure. In this section, we present the formulations of an MPC-based model for optimizing EVA capacity offerings, and RT power control of connected EVs to respond to regulation signals.

Recall that the offered regulation capacity of EVA is allowed to be changed on an hourly basis one time slot before the operating hour $K$. To tackle with the EVA dynamics, the proposed stochastic MPC scheme solves a two-stage stochastic programming problem at every time $K$ over the prediction horizon $H_K := \{K, K+1, K+2, \ldots, K+H\}$, yet only solutions for the current time $K$ are implemented. In the next time slot, EVA shall update the system information (e.g., the set of connected EVs, their required charging energy, forecasting information etc.) and re-do the optimizations. Such a formulation allows to include: (i) EVA operation model, (ii) predictions of the parameters affecting EVA's performance, (iii) constraints of regulation bids cleared in the market which should be delivered.

### A. Aggregated EV Model for Prediction

Based on theorem 1 and 2, we partition V1G and V2G vehicles into a finite number of "equivalent EVs", each representing a specific group of a particular arrival/departure time and charging flexibility index. As a result, we utilize a virtual EV set to evaluate available capacities of EV fleet. In the virtual set, there are a constant number of equivalent EVs(considering all possible realizations), each with a unique property (i.e. arrival/departure time, charging flexibility index and charging mode). Thereby, only the charging parameters of each virtual EV (i.e. required energy and maximal power rate) need to be predicted. Advantages of this predictive model are: i) the parking time of each virtual EV is known, so predicted parameters have been reduced; ii) the unknown parameters are relatively steady and small for prediction, which represent the cumulative charging characteristics of a specific EV group.

In order to optimize the economic scheduling of an EVA by MPC approach with $H$ look-ahead intervals at each time, all the future information in the prediction horizon is needed. We utilize the aggregated model proposed above to evaluate the available capacity of upcoming EVs within next time slot. Let $\Lambda_1/\Lambda_2$ be the virtual EV sets in V1G/V2G modes for prediction, and let $\Xi_1/\Xi_2$ denote the sets of all V1G/V2G EVs that have been already connected in the aggregator. We rely on existing approaches (e.g., time series, probabilistic forecasting and support vector regression etc.) to predict market prices of energy and regulation, as well as parameters in virtual EV sets. Details of these forecast methods can be found in [21]-[23] and will be omitted in this paper. In our stochastic MPC, a set of scenarios $\{s \in \Omega, \pi_s\}$ of energy prices $\lambda_\tau^{(s)}$, regulation prices $\mu_\tau^{(s)}$, and predicted EV vectors $[\mathbf{E}_\tau^{\Lambda_1/\Lambda_2,(s)}; \mathbf{P}_+^{\Lambda_1/\Lambda_2,(s)}]$, are generated.

### B. Rolling-horizon Optimization

The complete two-stage stochastic programming problem can be formulated as follows, which is presented in a matrix form:

$$\text{Min } Cost^{(s)} = \left( C_K + C_{K+1}^{(s)} + CTol^{(s)} + Pnlty^{(s)} \right), \quad (19)$$

where

$$C_K = \lambda_K \mathbf{1}^\mathbf{T} \left[ \mathbf{X}_K^{\Xi_1}; \mathbf{X}_K^{\Xi_2} \right] - (\lambda_K - \psi) \mathbf{1}^\mathbf{T} \mathbf{Y}_K^{\Xi_2}; \quad (20)$$

$$C_{K+1}^{(s)} = \lambda_{K+1}^{(s)} \mathbf{1}^\mathbf{T} \left[ \mathbf{X}_{K+1}^{\Xi_1}; \mathbf{X}_{K+1}^{\Xi_2}; \mathbf{X}_{K+1}^{\Lambda_1}; \mathbf{X}_{K+1}^{\Lambda_2} \right] \\ - \left(\lambda_{K+1}^{(s)} - \psi\right) \mathbf{1}^\mathbf{T} \left[ \mathbf{Y}_{K+1}^{\Xi_2}; \mathbf{Y}_{K+1}^{\Lambda_2} \right] - \mu_{K+1}^{(s)} R_{K+1}; \quad (21)$$

$$CTol^{(s)} = \sum_{\tau=K+2}^{K+H} \left\{ \begin{array}{l} \lambda_\tau^{(s)} \mathbf{1}^\mathbf{T} \left[ \mathbf{X}_\tau^{\Xi_1,(s)}; \mathbf{X}_\tau^{\Xi_2,(s)}; \mathbf{X}_\tau^{\Lambda_1,(s)}; \mathbf{X}_\tau^{\Lambda_2,(s)} \right] \\ - \left(\lambda_\tau^{(s)} - \psi\right) \mathbf{1}^\mathbf{T} \left[ \mathbf{Y}_\tau^{\Xi_2,(s)}; \mathbf{Y}_\tau^{\Lambda_2,(s)} \right] \\ - \mu_\tau^{(s)} \mathbf{1}^\mathbf{T} \left[ \mathbf{Z}_\tau^{\Xi_1,(s)}; \mathbf{Z}_\tau^{\Xi_2,(s)}; \mathbf{Z}_\tau^{\Lambda_1,(s)}; \mathbf{Z}_\tau^{\Lambda_2,(s)} \right] \end{array} \right\}; \quad (22)$$

$$Pnlty^{(s)} = \varphi \omega_K + \varphi' \omega_{K+1}^{(s)}. \quad (23)$$

Subject to
- Constraints for V1G vehicles:

$$\mathbf{Z}_K^{\Xi_1} \leq \mathbf{X}_K^{\Xi_1} \leq \mathbf{P}_+^{\Xi_1} - \mathbf{Z}_K^{\Xi_1}, \quad (24)$$

$$\mathbf{Z}_\tau^{\Xi_1,(s)} \leq \mathbf{X}_\tau^{\Xi_1,(s)} \leq \mathbf{P}_+^{\Xi_1} - \mathbf{Z}_\tau^{\Xi_1,(s)}, \forall s,\tau, \quad (25)$$

$$\mathbf{Z}_\tau^{\Lambda_1,(s)} \leq \mathbf{X}_\tau^{\Lambda_1,(s)} \leq \mathbf{P}_+^{\Lambda_1,(s)} - \mathbf{Z}_\tau^{\Lambda_1,(s)}, \forall s,\tau, \quad (26)$$

$$\mathbf{Z}_K^{\Xi_1}, \mathbf{Z}_\tau^{\Xi_1,(s)}, \mathbf{Z}_\tau^{\Lambda_1,(s)} \geq \mathbf{0}, \forall s,\tau, \quad (27)$$

$$\left[\mathbf{A}_\tau^{\Xi_1}; \mathbf{A}_\tau^{\Lambda_1}\right] \circ \left[\mathbf{X}_\tau^{\Xi_1,(s)}; \mathbf{X}_\tau^{\Lambda_1,(s)}\right] = \mathbf{0}, \forall s,\tau, \quad (28)$$

$$g(H,\Gamma) = \min\left\{1, \frac{H}{\Gamma}\right\}, \quad (29)$$

$$\mathbf{X}_K^{\Xi_1} + \sum_{\tau=K+1}^{K+H} \mathbf{X}_\tau^{\Xi_1,(s)} = \mathbf{E}_r^{\Xi_1} \circ \boldsymbol{g}\left(H, \boldsymbol{\Gamma}_K^{\Xi_1}\right), \quad (30)$$

$$\sum_{\tau=K+1}^{K+H} \mathbf{X}_\tau^{\Lambda_1,(s)} = \mathbf{E}_r^{\Lambda_1,(s)} \circ \boldsymbol{g}\left(H, \boldsymbol{\Gamma}_K^{\Lambda_1}\right); \quad (31)$$

- Constraints for V2G vehicles:

$$\mathbf{Z}_K^{\Xi_2} + \mathbf{X}_K^{\Xi_2} \leq \mathbf{P}_+^{\Xi_2}, \quad \mathbf{Z}_K^{\Xi_2} + \mathbf{Y}_K^{\Xi_2} \leq \mathbf{P}_+^{\Xi_2}, \quad (32)$$

$$\mathbf{Z}_\tau^{\Xi_2,(s)} + \mathbf{X}_\tau^{\Xi_2,(s)} \leq \mathbf{P}_+^{\Xi_2}, \mathbf{Z}_\tau^{\Xi_2,(s)} + \mathbf{Y}_\tau^{\Xi_2,(s)} \leq \mathbf{P}_+^{\Xi_2}, \forall s,\tau, \quad (33)$$

$$\mathbf{Z}_\tau^{\Lambda_2,(s)} + \mathbf{X}_\tau^{\Lambda_2,(s)} \leq \mathbf{P}_+^{\Lambda_2,(s)}, \mathbf{Z}_\tau^{\Lambda_2,(s)} + \mathbf{Y}_\tau^{\Lambda_2,(s)} \leq \mathbf{P}_+^{\Lambda_2,(s)}, \forall s,\tau, \quad (34)$$

$$\mathbf{X}_K^{\Xi_2}, \mathbf{Y}_K^{\Xi_2}, \mathbf{Z}_K^{\Xi_2}, \mathbf{X}_\tau^{\Xi_2/\Lambda_2,(s)}, \mathbf{Y}_\tau^{\Xi_2/\Lambda_2,(s)}, \mathbf{Z}_\tau^{\Xi_2/\Lambda_2,(s)} \geq \mathbf{0}, \forall s,\tau, \quad (35)$$

$$\left[\mathbf{A}_\tau^{\Xi_2}; \mathbf{A}_\tau^{\Lambda_2}\right] \circ \left[\mathbf{X}_\tau^{\Xi_2,(s)}; \mathbf{X}_\tau^{\Lambda_2,(s)}\right] = \mathbf{0}, \forall s,\tau, \quad (36)$$

$$\left[\mathbf{A}_\tau^{\Xi_2}; \mathbf{A}_\tau^{\Lambda_2}\right] \circ \left[\mathbf{Y}_\tau^{\Xi_2,(s)}; \mathbf{Y}_\tau^{\Lambda_2,(s)}\right] = \mathbf{0}, \forall s,\tau, \quad (37)$$

$$\left[\mathbf{A}_\tau^{\Xi_2}; \mathbf{A}_\tau^{\Lambda_2}\right] \circ \left[\mathbf{Z}_\tau^{\Xi_2,(s)}; \mathbf{Z}_\tau^{\Lambda_2,(s)}\right] = \mathbf{0}, \forall s,\tau, \quad (38)$$

$$\mathbf{E}_-^{\Xi_2} \leq \sum_\tau \left(\mathbf{X}_\tau^{\Xi_2,(s)} - \mathbf{Y}_\tau^{\Xi_2,(s)}\right) \leq \mathbf{E}_+^{\Xi_2}, \forall s,\tau, \quad (39)$$

$$\mathbf{E}_-^{\Lambda_2,(s)} \leq \sum_\tau \left(\mathbf{X}_\tau^{\Lambda_2,(s)} - \mathbf{Y}_\tau^{\Lambda_2,(s)}\right) \leq \mathbf{E}_+^{\Lambda_2,(s)}, \forall s,\tau, \quad (40)$$

$$\left(\mathbf{X}_K^{\Xi_2} - \mathbf{Y}_K^{\Xi_2}\right) + \sum_{\tau=K+1}^{K+H} \left(\mathbf{X}_\tau^{\Xi_2,(s)} - \mathbf{Y}_\tau^{\Xi_2,(s)}\right) = \mathbf{E}_r^{\Xi_2} \circ \boldsymbol{g}\left(H, \boldsymbol{\Gamma}_K^{\Xi_2}\right), \quad (41)$$

$$\sum_{\tau=K+1}^{K+H} \left(\mathbf{X}_\tau^{\Lambda_2,(s)} - \mathbf{Y}_\tau^{\Lambda_2,(s)}\right) = \mathbf{E}_r^{\Lambda_2,(s)} \circ \boldsymbol{g}\left(H, \boldsymbol{\Gamma}_K^{\Lambda_2}\right); \quad (42)$$

- Constraints for EVA cleared regulation capacities:

$$\mathbf{1}^\mathrm{T}\left[\mathbf{Z}_K^{\Xi_1}; \mathbf{Z}_K^{\Xi_2}\right] \geq R_K - \omega_K, \quad (43)$$

$$\mathbf{1}^\mathrm{T}\left[\mathbf{Z}_\tau^{\Xi_1,(s)}; \mathbf{Z}_\tau^{\Xi_2,(s)}; \mathbf{Z}_\tau^{\Lambda_1,(s)}; \mathbf{Z}_\tau^{\Lambda_2,(s)}\right] \geq R_{K+1} - \omega_{K+1}^{(s)}, \quad (44)$$

$$\omega_K, \omega_{K+1}^{(s)} \geq 0, \quad \forall s. \quad (45)$$

The objective function (19) is to minimize the total cost within $H$ look-ahead intervals, considering both energy and regulation markets. The first term (20) defines the EVA energy cost at the operating hour $K$, which considers the battery degradation cost caused by discharging. Note that the regulation capacity of current time slot $K$ has been cleared at time $K$-1; thus is not included. The energy and regulation cost at the next time slot $K$+1, and the total costs from time slot $K$+2 to time slot $K$+$H$, are defined separately in (21) and (22). The last term (23) defines the penalty of unfulfilled regulation capacities, which reflects the regulation performance of the EVA. The penalty factor of a unit unfulfilled regulation capacity at time $K$ or at time $K$+1 in scenario $s$ is denoted by $\omega_K$, $\omega_{K+1}^{(s)}$, respectively.

Constraints (24)-(31) model operational constraints of connected and upcoming V1G sets. In (24)-(26), charging power under the scheduled POP and committed regulation capacity must be within 0 and maximum power rating. Regulation capacities always hold non-negative in (27). The sign ∘ denotes the element-wise product. Constraint (28) defines the charging availability, which enforces the POPs during the unconnected periods to be 0; correspondingly, the regulation capacities will be zero. The vector function expressed in (29), denotes the ratio of required energy within $H$ window width to that before the departure deadline. It helps to build the end constraints in the MPC scheme. To make sure the charging requests of EV owners, the charging POPs within $H$ look-ahead intervals are constrained to reach the target energy in (30)-(31).

Constraints (32)-(42) model operational constraints of connected and upcoming V2G sets. In (32)-(34), the available regulation capacities are constrained by the charging POPs, discharging POPs and the maximum power rating. Constraint (35) define the non-negativity of V2G EVs' operation variables. Constraint (36)-(38) define the availability of the EVs to the aggregator, which ensures all variables of an EV (i.e., charging/discharging POPs, and regulation capacities) to be zero after its departure. In (39)-(40), the energy level of EV battery is constrained to remain in its permissible range. Constraints (41)-(42) define expected energy demands of EVs during the prediction horizon to satisfy their charging requests.

Allocations of the EVA cleared regulation capacities are modeled in (43)-(45). In (43), the cleared regulation capacity from the last market period should be fulfilled by connected EVs at the current time $K$. In (44), the optimal regulation capacity at the next time slot is estimated based on the available EV fleets to the aggregator. In (45), variables of unfulfilled regulation must stay non-negative.

In the optimization problem above, the first-stage (here and now) variables are the scheduled POPs and regulation capacities of all connected EVs in both set $\Xi_1$ and set $\Xi_2$ at the operating hour $K$, $\mathbf{X}_K^{\Xi_1/\Xi_2}, \mathbf{Y}_K^{\Xi_2}, \mathbf{Z}_K^{\Xi_1/\Xi_2}$, the unfulfilled regulation capacity at time $K$, $\omega_K$, and the regulation capacity of the next time slot offered to the market before the operating hour, $R_{K+1}$; while the second-stage (wait and see) decision variables are $\mathbf{X}_\tau^{\Xi/\Lambda,(s)}, \mathbf{Y}_\tau^{\Xi_2/\Lambda_2,(s)}, \mathbf{Z}_\tau^{\Xi/\Lambda,(s)}$ ($\tau$=$K$+1,...,$K$+$H$) and $\omega_{K+1}^{(s)}$, where $s$ represents the scenario index.

*C. Modeling Risk-Averseness*

To trade off the expected cost and risks in the two-stage stochastic optimization problem (19)-(45), CVaR is introduced in the objective function. For a given confidence level denoted by $\alpha \in (0,1)$, value-at-risk (VaR) is defined as the cost in the $(1-\alpha)$ fraction of worst-case outcomes, which is denoted by *Var* in this optimization. CVaR denotes the expected value of *Var*, which can be mathematically defined as:

$$CVaR_{\alpha,s} = \mathrm{Min}\left\{Var + \frac{1}{1-\alpha}\pi_s\left\{\max\left\{Cost^{(s)} - Var, 0\right\}\right\}\right\}. \quad (46)$$

To formally incorporate CVaR into the optimization, associate probabilities $\pi_s$ to the predicted scenarios and let $\alpha' = (1-\alpha)^{-1}$, the risk-sensitive EVA problem can be presented as:

$$\begin{aligned}
\mathrm{Min} \quad & \left\{Var + \alpha'\sum_s \pi_s v^{(s)}\right\}, \\
\text{subject to} \quad & v^{(s)} \geq 0, \\
& v^{(s)} \geq C_K + C_{K+1}^{(s)} + CTol^{(s)} + Pnlty^{(s)} - Var, \\
& \text{relations } (20)-(45) \\
& \text{for each } s \in \Omega.
\end{aligned} \quad (47)$$

The linearization of (46) is achieved by adding new auxiliary variables $v^{(s)}$. Taking different value of $\alpha \in [0,1]$ can tune the EVA tolerance to higher costs. Choosing α equal to zero, the aggregator will treat all scenarios equally and minimizes the expected cost; while as α increases, the EVA will weigh scenarios where the cost is higher more heavily.

*D. Real-Time Dispatching Control*

The dynamic EVA problem (47) is a linear programming problem which can be solved efficiently at the start of each operating hour. In particular, optimal solutions to the first-stage variables can be denoted as $\mathbf{X}_K^{\Xi_1/\Xi_2*}, \mathbf{Y}_K^{\Xi_2*}, \mathbf{Z}_K^{\Xi_1/\Xi_2*}, \omega_K^*$, and $R_{K+1}^*$.

Within the time interval of the current operating hour, constant POPs and committed regulation tasks can be allocated by the aggregator to connected EVs as:

$$\mathbf{P}_K^{\Xi_1} = \mathbf{X}_K^{\Xi_1*}, \mathbf{P}_K^{\Xi_2} = \mathbf{X}_K^{\Xi_2*} - \mathbf{Y}_K^{\Xi_2*}, \quad (48)$$

$$\mathbf{R}_K^{\Xi_1/\Xi_2} = \begin{cases} R_K \times \dfrac{\mathbf{Z}_K^{\Xi_1/\Xi_2*}}{\mathbf{1}^T\left[\mathbf{Z}_K^{\Xi_1*}; \mathbf{Z}_K^{\Xi_2*}\right]}, & \omega_K^* = 0, \\ \mathbf{Z}_K^{\Xi_1/\Xi_2*}, & \omega_K^* > 0, \end{cases} \quad (49)$$

where $\mathbf{P}_K^{\Xi_1/\Xi_2}$, $\mathbf{R}_K^{\Xi_1/\Xi_2}$ denote the scheduled POPs and regulation capacities of plug-in EVs at the operating time slot, respectively. In (49), the cleared regulation task is assigned proportionally within the available regulation capability.

Moreover, charging power of connected EVs may change around the constant scheduled POPs in the actual regulation delivery. The ISO will broadcast capacity normalized AGC signals to related resources. In PJM, the RegD signals are broadcast every 2 second. Then, the EVA must respond to by modulating power profiles of plug-in EVs, as determined by:

$$\mathbf{S}_{\gamma,K}^{\Xi_1/\Xi_2} = \mathbf{P}_K^{\Xi_1/\Xi_2} - reg_D(\gamma, K) \times \mathbf{R}_K^{\Xi_1/\Xi_2}, \quad (50)$$

where $reg_D(\gamma, K) \in [-1,1]$ denotes the RegD signals sent by ISO: $reg_D(\gamma, K) > 0$, means the up-regulation service is required and vice versa. $\mathbf{S}_{\gamma,K}^{\Xi_1/\Xi_2}$ denotes the RT power of connected EV set $\Xi_1/\Xi_2$ at time $K$, which changes with the 2-sec varying RegD signal.

From (50), we can see that energy exchange of EVs coming from both the scheduled charging/discharging and regulation provisions. To ensure charging requests from EVs, the remained parking time, required energy, and upper/lower bounded energy for the connected EV sets need to be corrected at the end of the operating hour. Let $Reg_D^K = \sum_\gamma reg_D(\gamma, K)$ denotes the hourly cumulated regulation signals at time $K$. Given the operational decisions and regulation signals, their values adjust to be:

$$\boldsymbol{\Gamma}_{K+1}^{\Xi_1/\Xi_2} = \boldsymbol{\Gamma}_K^{\Xi_1/\Xi_2} - \mathbf{1}. \quad (51)$$

$$\mathbf{E}_r^{\Xi_1/\Xi_2} := \mathbf{E}_r^{\Xi_1/\Xi_2} - \left[\mathbf{P}_K^{\Xi_1/\Xi_2} - Reg_D(K) \times \mathbf{R}_K^{\Xi_1/\Xi_2}\right], \quad (52)$$

$$\mathbf{E}_{+/-}^{\Xi_2} := \mathbf{E}_{+/-}^{\Xi_2} - \left[\mathbf{P}_K^{\Xi_2} - Reg_D(K) \times \mathbf{P}_K^{\Xi_2}\right]. \quad (53)$$

To capture the EVA dynamic, MPC-based scheme is adopted to manage power dispatching of connected EVs in RT operation. During each operating hour, $K$, it holds as follows:
1) At start of the operating hour, the two-stage stochastic programming problem (47) is solved; then the optimal regulation capacity $R_{K+1}^*$ is submitted to ISO by the EVA.
2) At the time interval, the EVA modules the RT power of connected EVs as (50) varying with RegD signals.
3) At the end of operating hour, parameters of plug-in EV sets are coordinated as shown in (51)-(53); then the EVA modify the connected EV sets by adding the new comers and deleting leaving ones; moreover, all predicted scenarios are updated based on the latest observations.
4) The MPC time horizon recedes by a step, $K \leftarrow K+1$, for the next time slot.

TABLE III
CHARACTERISTIC PARAMETERS OF DIFFERENT TYPES OF EVS

| Types | Num. | Arrival time(h) | Departure time(h) | Arrival SoC | Target SoC | Rated energy capacity | Maximal power rate |
|---|---|---|---|---|---|---|---|
| I | 1200 | 16-23 | 6-13 | U, (0.2,0.4) | U, (0.7,0.9) | U,(25,45) [kWh] | U,(5,8) [kWh] |
| II | 400 | 0-7 | 14-21 | | | | |
| III | 400 | 8-15 | 22-5 | | | | |

*U: uniform distribution; and (α,β) is the variation range.

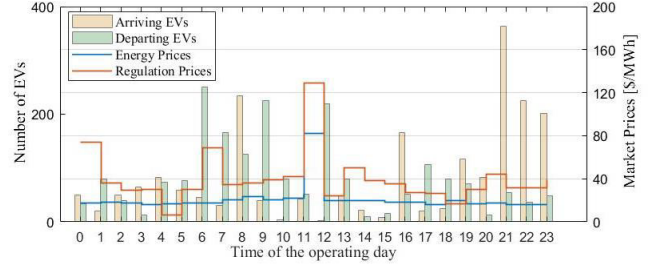

Fig. 4. The dynamic EVA during the operation day.

V. NUMERICAL EXPERIMENTS

We investigate the proposed method in a 2000-EV test system during a particular day to validate its performance, where EVA participates in RT markets with its available EVs in a rolling horizon manner. All the computations are solved in MATLAB embedded with CPLEX on the laptop with 2.0GHz Intel(R) Core(TM) i7-8565U microprocessor and 8GB RAM.

*A. Parameters Setting*

We assume that there are 3 types of driving patterns among the 2,000 heterogeneous EVs, whose characteristic parameters are generated based on the distribution in Table III. Only slow charging is considered to explore the flexibility of EVs. Moreover, we assume 50% of the EVs in each type are in V1G mode; while the others are in V2G mode. The minimum and maximum SoC are limited to 15% and 90%. The battery degradation price $\psi$ is set as 50/$MWh. Historical market clearing prices and RegD signals produced by PJM for a particular day (Nov. 25, 2020) are used, and all data is available to the public [24]-[25]. The EVA dynamics and market clearing prices during operating day are shown in Fig.4. In the MPC-based problem, we consider that the window width $H$ =8 and the confidence level of CVaR $\alpha$ =0.2. There are 100 scenarios generated by Monte Carlo method, where all predicted errors are assumed to obey the Gaussian distributions. The standard covariances of market price $\varepsilon_p$ and upcoming EV demand (both the energy and power rating) $\varepsilon_{EV}$ are set to be 3$/MWh and 2kWh. To model the accuracy of prediction evolving with time, covariances of market prices over $\mathcal{H}_K := \{K+1, K+2, \ldots, K+H\}$ are assumed to be $\{\varepsilon_p, 2*\varepsilon_p, \ldots, H*\varepsilon_p\}$. In addition, we assume that the penalty factors $\varphi$ and $\varphi'$ are set to be 130 and 40 $/MWh, respectively. Here, the value of $\varphi$ is larger than the maximal regulation price in the day, therefore, EVA will try to honor the cleared regulation bids within its available capacities in the RT delivery.

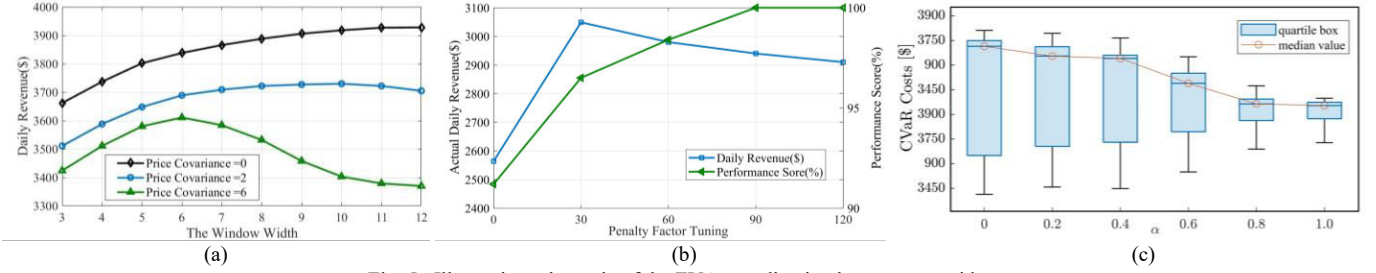

Fig. 5. Illustrative schematic of the EVA coordination between two sides.
(a) Regret analysis in the MPC scheme. (b) Performance score and actual revenue versus penalty factor. (c) The CVaR costs versus confidence levels.

TABLE IV
SUMMARY OF DIFFERENT STRATEGIES

| Strategy | | Energy Cost $ | Degradation Cost $ | Regulation Payment $ | Daily Revenue $ |
|---|---|---|---|---|---|
| Immediate Charging | | 697.3 | 0 | 0 | -697.3 |
| Smart Charging | [V1G] | 589.8 | 0 | 0 | -589.8 |
| | [V2G] | 149.6 | 339.0 | | -488.5 |
| Proposed MPC [Mix] | | 821.3 | 0 | 3867.1 | 3045.8 |
| Ideal MPC | [V1G] | 929.5 | | 1792.6 | 863.1 |
| | [V2G] | 706.3 | 0 | 6157.4 | 5451.1 |
| | [Mix] | 834.7 | | 4025.2 | 3190.5 |
| Robust MPC [Mix] | | 819.8 | 0 | 3639.4 | 2819.6 |

TABLE V
WORST SOC DEVIATIONS OF EVS AT DEPARTURE

| Charging mode | Unidirectional charging(V1G) | Bidirectional charging(V2G) |
|---|---|---|
| Worst SoC deviation (%) | 0.91 | 1.57 |

## B. Results for the Base Case

Economic performance of the proposed method is compared with different strategies. Immediate charging means EVs start to charging at their maximum power rating as soon as they arrive at the aggregator. The smart charging strategy is only adopted to optimize energy decisions but ignores regulation market participation. The following three strategies consider both energy and regulation markets. Ideal MPC, where all future data is perfectly predicted, is adopted to shows the theoretically optimal result. However this strategy is not implementable in practice. The robust MPC strategy adopts our proposed method but ignore the upcoming EV demands in the prediction part. Furthermore, we consider three charging modes here: a) V1G mode, where unidirectional charging is utilized by the total 2000-EVs; b) V2G mode, where bidirectional charging is utilized by the total 2000-EVs; c) Mix mode, with 50% of the 2000-EVs in V1G mode; and the rest in V2G mode, as set in the EVA system. The summary of their energy cost, regulation payments, and daily revenue after one day of operation are listed in Table IV.

From the results, we find that V2G vehicles in the smart charging strategy have performed energy arbitrage to minimize the total cost, while V2G vehicles in ideal MPC strategy have zero degradation cost. This is because the biggest gap of energy prices in operating day is greater than the battery degradation price. Moreover, the relationship $\{\mu_\tau+\psi>\lambda_\tau: \forall \tau\}$ holds, as shown in lemma 2, V2G vehicles will keep the POP in charging mode to retain more regulation capacities for profit. Moreover, one downside of strategies considering regulation participation is the higher energy cost for V2G vehicles in comparison to strategies only considering energy markets, since they try to optimize gross revenue by exploiting the difference between energy and regulation prices. In the ideal MPC strategy, the regulation revenue in V2G mode is three times as much as that in V1G mode. This is because the total regulation capacities offered by a V1G EV during its parking periods, must be limited to its required energy, while a V2G EV can provide regulation service by its discharging. Compared to our proposed MPC, the robust MPC strategy results in an 8% reduction in daily revenue, which indicates that the strategy does not take full advantage of the EVA flexibility, as the upcoming EV demand is ignored. The comparisons validate the economic performance of the proposed method.

In addition, we study the deviation between the actual SoC and target SoC of all EVs at their departure. This is because from the EV owners' perspective, whether the target SoC is reached at the time of departure is the primary concern. The worst SoC deviation results are given in the Table V, where the worst value among V2G EVs (1.57%) is almost twice of that among V1G EVs (0.91%). In the MPC scheme, this SoC deviation is mainly caused by non-energy-neutral regulation delivery in the last parking time slot. And it can be reduced by shortening time interval of the optimization. For both charging modes, these SoC deviations to be acceptable to EV owners in the practical applications.

## C. Impact of the Window Width H

There are two factors related with the window width $H$ that may affect the EVA economic performance: the proportional energy function in (29) which helps to build end constraints in the MPC; and prediction errors of market prices over the horizon $\mathcal{H}_K := \{K+1, K+2, …, K+H\}$. To demonstrate the relationship, perfect forecast of future EV demands and the zero hourly RegD signals are adopted here. Fig. 5(a) shows the EVA daily performances versus window widths $H$ under different $\varepsilon_p$ value.

We can see that with the perfect prediction of market prices, the optimality gap will obviously decrease with a larger $H$ because more foreseen prices are used in the optimization. In addition, the daily revenue under $\varepsilon_p = 0$ here is much larger than the ideal MPC strategy in Table IV. This is because the energy deviation caused by non-zero hourly RegD signals may lead to an unexpected energy cost and a decrease in subsequent regulation capacity bids for V1G EVs. When $\varepsilon_p$ exceeds a certain value, the EVA performance will first increase and then decrease as $H$ increases. Under such cases, as it can be observed, the optimal window width $H$ decreases as the $\varepsilon_p$ value increases.

*D. Impact of the Penalty Factor $\varphi'$*

The parameter $\varphi'$ sets the penalty factor of a unit unfulfilled regulation capacity of the next time slot, and its value affects the bidding decisions of EVA in regulation markets. Under the performance-based regulation compensation scheme, performance score is introduced to reflect the accuracy of regulation delivery with issued signals. Considering that it is difficult to calculate performance scores in the optimization when a resource fails to respond to regulation signals, we build the unfulfilled regulation penalty term (23) in the objective to punish the over-capacity regulation offer. After the operation, performance score and actual revenue can be calculated, and the method in [11] is adopted here.

We study the impact of penalty factor $\varphi'$ on the performance score and actual daily revenue. Results after the EVA one-day operation under different $\varphi'$, i.e., taking values from 0 to 120, are shown in Fig.5(b). It can be seen that the optimal value $\varphi'$ seems to be 30\$/MWh, and the average regulation price in the operating day is 39\$/MWh. When $\varphi'$ increases, the probability of over-capacity regulation bids decreases; thus, the performance score will increase with a more conservative regulation offer. Moreover, values of the actual daily revenue demonstrate the trade-off between the regulation capacity bids and its performance score.

*E. Risk Aversion*

To test the risk aversion of our proposed method, we study the impact of different confidence levels of CVaR on the EVA daily revenue. We assume he zero hourly RegD signals here, and the uncertainties of market prices and upcoming EV demands will affect the EVA performance due to the unknown future realization during the RT operation. Firstly, we generated 50 different situations. And in each situation, there is a defined 100-scenario at any given time slot. At any confidence level, we run all the situations, and CVaR results are plotted as the boxchart in Fig. 5(c): the median daily revenue is shown as the line inside the box, and we can find that the median daily revenue keeps dropping as $\alpha$ increases; and the lower and upper quartiles are shown as the bottom and top edges of the box, which confirms the risk-averseness of the EVA.

## VI. Conclusion

This paper proposes an MPC-based operation methodology for the EVA participating in RT energy and regulation markets. Both the market bidding strategy and RT charging allocation mechanism are incorporated in our formulation. A CVaR based two-stage stochastic programming is developed to address the uncertainties of upcoming EVs as well as energy and regulation market prices. Simulations prove that the proposed rolling-horizon operation framework can effectively achieve the great delivery of cleared regulation services and the charging requests from EV owners. Numerical results also show the profitable revenue from the RT electricity markets, which opens up the opportunity for broader EVA applications.

## References

[1] IEA. Global EV Outlook 2020, 2020. [Online]. Available: http://www.https://www.iea.org/reports/global-ev-outlook-2020.
[2] Q. Chen, N. Liu, C. Hu, L. Wang and J. Zhang, "Autonomous Energy Management Strategy for Solid-State Transformer to Integrate PV-Assisted EV Charging Station Participating in Ancillary Service," in IEEE Transactions on Industrial Informatics, vol. 13, no. 1, pp. 258-269, Feb. 2017, doi: 10.1109/TII.2016.2626302.
[3] K. Knezović, S. Martinenas, P. B. Andersen, A. Zecchino and M. Marinelli, "Enhancing the Role of Electric Vehicles in the Power Grid: Field Validation of Multiple Ancillary Services," in IEEE Transactions on Transportation Electrification, vol. 3, no. 1, pp. 201-209, March 2017, doi: 10.1109/TTE.2016.2616864.
[4] Y. Yu, O. S. Nduka and B. C. Pal, "Smart Control of an Electric Vehicle for Ancillary Service in DC Microgrid," in IEEE Access, vol. 8, pp. 197222-197235, 2020, doi: 10.1109/ACCESS.2020.3034496.
[5] A. Fachechi, L. Mainetti, L. Palano, L. Patrono, M. L. Stefanizzi, R. Vergallo, P. Chu, and R. Gadh, "A new vehicle-to-grid system for battery charging exploiting IoT protocols," in Proc. IEEE Int. Conf. Ind. Technol. (ICIT), Mar. 2015, pp. 2154–2159.
[6] PJM Manual 11: Energy & Ancillary Services Market Operations Revision. [Online]. Available: http://www.pjm.com.
[7] M. Wang, Y. Mu, Q. Shi, H. Jia and F. Li, "Electric Vehicle Aggregator Modeling and Control for Frequency Regulation Considering Progressive State Recovery," in IEEE Transactions on Smart Grid, vol. 11, no. 5, pp. 4176-4189, Sept. 2020, doi: 10.1109/TSG.2020.2981843.
[8] F. L. Müller, J. Szabó, O. Sundström and J. Lygeros, "Aggregation and Disaggregation of Energetic Flexibility From Distributed Energy Resources," in IEEE Transactions on Smart Grid, vol. 10, no. 2, pp. 1205-1214, March 2019, doi: 10.1109/TSG.2017.2761439.
[9] H. Zhang, Z. Hu, Z. Xu and Y. Song, "Evaluation of Achievable Vehicle-to-Grid Capacity Using Aggregate PEV Model," in IEEE Transactions on Power Systems, vol. 32, no. 1, pp. 784-794, Jan. 2017.
[10] T. Long, Q. -S. Jia, G. Wang and Y. Yang, "Efficient Real-Time EV Charging Scheduling via Ordinal Optimization," in IEEE Transactions on Smart Grid, vol. 12, no. 5, pp. 4029-4038, Sept. 2021, doi: 10.1109/TSG.2021.3078445.
[11] H. Zhang, Z. Hu, E. Munsing, S. J. Moura and Y. Song, "Data-Driven Chance-Constrained Regulation Capacity Offering for Distributed Energy Resources," in IEEE Transactions on Smart Grid, vol. 10, no. 3, pp. 2713-2725, May 2019.
[12] X. Ke, D. Wu and N. Lu, "A Real-Time Greedy-Index Dispatching Policy for Using PEVs to Provide Frequency Regulation Service," in IEEE Transactions on Smart Grid, vol. 10, no. 1, pp. 864-877, Jan. 2019.
[13] H. Liu, J. Qi, J. Wang, P. Li, C. Li and H. Wei, "EV Dispatch Control for Supplementary Frequency Regulation Considering the Expectation of EV Owners," in IEEE Transactions on Smart Grid, vol. 9, no. 4, pp. 3763-3772, July 2018.
[14] S. I. Vagropoulos, D. K. Kyriazidis and A. G. Bakirtzis, "Real-Time Charging Management Framework for Electric Vehicle Aggregators in a Market Environment," in IEEE Transactions on Smart Grid, vol. 7, no. 2, pp. 948-957, March 2016.
[15] H. Nakano et al., "Aggregation of V2H Systems to Participate in Regulation Market," in IEEE Transactions on Automation Science and Engineering, vol. 18, no. 2, pp. 668-680, April 2021, doi: 10.1109/TASE.2020.3001060.
[16] G. Wenzel, M. Negrete-Pincetic, D. E. Olivares, J. MacDonald and D. S. Callaway, "Real-Time Charging Strategies for an Electric Vehicle Aggregator to Provide Ancillary Services," in IEEE Transactions on Smart Grid, vol. 9, no. 5, pp. 5141-5151, Sept. 2018.
[17] S. Cai and R. Matsuhashi, "Model Predictive Control for EV Aggregators Participating in System Frequency Regulation Market," in IEEE Access, vol. 9, pp. 80763-80771, 2021.
[18] R. Kumar, M. J. Wenzel, M. J. Ellis, M. N. ElBsat, K. H. Drees and V. M. Zavala, "A Stochastic Model Predictive Control Framework for Stationary Battery Systems," in IEEE Transactions on Power Systems, vol. 33, no. 4, pp. 4397-4406, July 2018, doi: 10.1109/TPWRS.2017.2789118.
[19] D Fernández-Muoz, JI Pérez-Díaz, I Guisández, et al. Fast frequency control ancillary services: An international review[J]. Renewable and Sustainable Energy Reviews, 2020, 120:1-17.
[20] M Elliott, LG Swan, M Dubarry, G Baure. "Degradation of electric vehicle lithium-on batteries in electricity grid services," in Journal of Energy Storage, 2020;32:101873. https://doi.org/10.1016/j.est.2020.101873.

[21] H. Chitsaz, P. Zamani-Dehkordi, H. Zareipour and P. P. Parikh, "Electricity Price Forecasting for Operational Scheduling of Behind-the-Meter Storage Systems," in IEEE Transactions on Smart Grid, vol. 9, no. 6, pp. 6612-6622, Nov. 2018, doi: 10.1109/TSG.2017.2717282.
[22] Z. Zhang and M. Wu, "Predicting Real-Time Locational Marginal Prices: A GAN-Based Approach," in IEEE Transactions on Power Systems, vol. 37, no. 2, pp. 1286-1296, March 2022, doi: 10.1109/TPWRS.2021.3106263.
[23] X. Zhang, K. W. Chan, H. Li, H. Wang, J. Qiu and G. Wang, "Deep-Learning-Based Probabilistic Forecasting of Electric Vehicle Charging Load With a Novel Queuing Model," in IEEE Transactions on Cybernetics, vol. 51, no. 6, pp. 3157-3170, June 2021, doi: 10.1109/TCYB.2020.2975134.
[24] PJM. "Data Miner 2 Tools for Public Data" [Online]. Available: https://dataminer2.pjm.com/list.
[25] PJM, "Ancillary Services," [Online]. Available: https://www.pjm.com/markets-and-operations/ancillary-service.aspx.

APPENDIX

*A. Proof of Lemma 1*

We study the optimal scheduling of a V1G EV $i$, which can be obtained from the problem (12), denoted by $[(x_{i,\tau}^*, z_{i,\tau}^*), \forall \tau]$. Firstly, we make a transformation to prove $|x_{i,\tau}^* - p_i^{max}/2| = |z_{i,\tau}^* - p_i^{max}/2|$. Assume that parking horizon of EV $i$ is denoted as $\mathcal{T}_i := \{1,\cdots,T_i\}$. Name $w_{i,\tau} = x_{i,\tau} - p_i^{max}/2$; $s_{i,\tau} = p_i^{max}/2 - z_{i,\tau}$. Then the optimization of EV $i$ in (12) can be rewritten as,

$$\text{Min} \sum_{\tau=1}^{T_i} \lambda_\tau w_{i,\tau} + \mu_\tau s_{i,\tau} + \frac{p_i^{max}}{2}(\lambda_\tau - \mu_\tau).$$

Subject to:

$$\sum_{\tau=1}^{T_i} w_{i,\tau} = e_i^r - \frac{p_i^{max}}{2} \times T_i, \quad (58)$$

$$|w_{i,\tau}| \le s_{i,\tau},$$

$$0 \le s_{i,\tau} \le \frac{p_i^{max}}{2}.$$

Note that, equation $|w_{i,\tau}^*| = s_{i,\tau}^*$ holds for any $\tau \in \mathcal{T}_i$ in (58). Because regulation market prices $\mu_\tau$ will always be larger than zero in the objective function. Therefore, we can eliminate $s_{i,\tau}$ by using $|w_{i,\tau}|$ to replace it while assuring the optimality of the problem. Moreover, to simplify the parameters, we name $v_{i,\tau} = 2 \times w_{i,\tau} / p_i^{max}$ as well as ignore the constant term and constant factor in the objective function. Then, we have,

$$\text{Min}_{v_{i,\tau}} \quad \mathcal{V} = \sum_{\tau=1}^{T_i} \lambda_\tau v_{i,\tau} + \mu_\tau |v_{i,\tau}|.$$

Subject to:

$$\sum_{\tau=1}^{T_i} v_{i,\tau} = e_i^r \times \frac{2}{p_i^{max}} - T_i, \quad (\omega) \quad (59)$$

$$-1 \le v_{i,\tau}, \quad (\rho_\tau)$$

$$v_{i,\tau} \le 1. \quad (\theta_\tau)$$

One can write the Karush-Kuhn-Tucker conditions for the minimization problem in (59) as

$$\sum_{\tau=1}^{T_i} v_{i,\tau} = e_i^r \times \frac{2}{p_i^{max}} - T_i = E_i, \quad (60a)$$

$$-1 \le v_{i,\tau} \quad \forall \tau \in \mathcal{T}_i \quad (60b)$$

$$v_{i,\tau} \le 1 \quad \forall \tau \in \mathcal{T}_i \quad (60c)$$

$$\frac{\partial \mathcal{V}}{\partial v_{i,\tau}} - \omega - \rho_\tau + \theta_\tau = 0 \quad \forall \tau \in \mathcal{T}_i \quad (60d)$$

$$\omega \left( \sum_{\tau=1}^{T_i} v_{i,\tau} - E_i \right) = 0, \quad (60e)$$

$$\rho_\tau (v_{i,\tau} + 1) = 0 \quad \forall \tau \in \mathcal{T}_i \quad (60f)$$

$$\theta_\tau (v_{i,\tau} - 1) = 0 \quad \forall \tau \in \mathcal{T}_i \quad (60g)$$

$$\rho_\tau \ge 0, \theta_\tau \ge 0 \quad \forall \tau \in \mathcal{T}_i \quad (60h)$$

Since the problem (59) only has linear constraints, and the subgradient of nonlinear objective function can be solved as,

$$\frac{\partial \mathcal{V}}{\partial v_{i,\tau}} = \begin{cases} \lambda_\tau - \mu_\tau & \text{when } v_{i,\tau} < 0 \\ [\lambda_\tau - \mu_\tau, \lambda_\tau + \mu_\tau] & \text{when } v_{i,\tau} = 0 \\ \lambda_\tau + \mu_\tau & \text{when } v_{i,\tau} > 0 \end{cases} \quad (61)$$

Note that the equation $\{\rho_\tau * \theta_\tau = 0: \forall \tau \in \mathcal{T}_i\}$ holds, because the two kinds of constraints will not be active simultaneously as seen from KKT conditions (60f) and (60g). $\omega$ is the optimal Lagrangian multiplier corresponding to the equality constraint in (59). We separate the analysis into four cases as follows,

1) for all time slots where $\lambda_\tau - \mu_\tau > \omega$, we have $\frac{\partial \mathcal{V}}{\partial v_{i,\tau}} - \omega = \rho_\tau - \theta_\tau > 0$. Then, $\rho_\tau > 0, \theta_\tau = 0$. From (60f), we have:

$$v_{i,\tau}^* = -1; \Rightarrow x_{i,\tau}^* = 0, z_{i,\tau}^* = 0. \quad (62a)$$

2) for all time slots where $\lambda_\tau + \mu_\tau < \omega$, we have $\frac{\partial \mathcal{V}}{\partial v_{i,\tau}} - \omega = \rho_\tau - \theta_\tau < 0$. Then, $\rho_\tau = 0, \theta_\tau > 0$. From (60g), we have:

$$v_{i,\tau}^* = 1; \Rightarrow x_{i,\tau}^* = p_i^{max}, z_{i,\tau}^* = 0. \quad (62b)$$

3) for all time slots where $\lambda_\tau - \mu_\tau < \omega < \lambda_\tau + \mu_\tau$, we have $v_{i,\tau}^* = 0; \Rightarrow x_{i,\tau}^* = p_i^{max}/2, z_{i,\tau}^* = p_i^{max}/2$. Otherwise, if $v_{i,\tau}^* > 0$; $\frac{\partial \mathcal{V}}{\partial v_{i,\tau}} - \omega > 0$. Then, we have $\rho_\tau > 0, \theta_\tau = 0$. $v_{i,\tau}^* = -1$, which is paradoxical; Else if $v_{i,\tau}^* < 0$; $\frac{\partial \mathcal{V}}{\partial v_{i,\tau}} - \omega < 0$. Then, we have $\rho_\tau = 0, \theta_\tau > 0$. $v_{i,\tau}^* = 1$, which is also paradoxical. To conclude, in this case, we have

$$v_{i,\tau}^* = 0; \Rightarrow x_{i,\tau}^* = p_i^{max}/2, z_{i,\tau}^* = p_i^{max}/2. \quad (62c)$$

4) for all time slots where $\lambda_\tau - \mu_\tau = \omega$, we have $v_{i,\tau}^* \in [-1,0]$; for all time slots where $\lambda_\tau + \mu_\tau = \omega$, we have $v_{i,\tau}^* \in [0,1]$. If there exists two time slots $(\alpha,\beta)$ in which $\lambda_\alpha - \mu_\alpha = \omega = \lambda_\beta - \mu_\beta$, or $\lambda_\alpha + \mu_\alpha = \omega = \lambda_\beta - \mu_\beta$, or $\lambda_\alpha + \mu_\alpha = \omega = \lambda_\beta + \mu_\beta$, we can find $v_{i,\alpha}^* + v_{i,\beta}^* = v_{i,\alpha}'^* + v_{i,\beta}'^*$, and both of them are optimal solutions. Therefore, we could always leave at most one time slot $v_{i,\chi}^*$ to be non-integer, and $\chi$ is defined as the marginal slot of the ith EV. ∎

*B. Proof of Theorem 1*

Note that all EVs in the set (where $i=1,\cdots,N$) have the same $t_i^a, t_i^d, F_i^{V1G}$ values. For simplicity, we assume that the common parking horizon is denoted as $\mathcal{T} := \{1,\cdots,T\}$ and the same charging flexibility index is denoted as $F$.

Firstly, we sort $2T+1$ elements of set $\{\lambda_\tau \pm \mu_\tau, -\infty\}$, $\forall \tau \in \mathcal{T}$ to get a new ascending ordered set $\{O_{(1)}, \ldots, O_{(2T+1)}\}$. Next, we prove $\omega = O_{(F+1)}$ is the solution to the KKT conditions in (60a) – (60h).

From discussions shown in (62), we can group $T$-1 parking time slots (except marginal time slot $\chi$) into three cases:

1) There are $a$ time slots, where $v_{i,\tau}^* = -1$, and $\lambda_\tau - \mu_\tau \geq \omega$;
2) There are $b$ time slots, where $v_{i,\tau}^* = 1$, and $\lambda_\tau + \mu_\tau \leq \omega$;
3) There are $c$ time slots, where $v_{i,\tau}^* = 0$, and $\lambda_\tau - \mu_\tau \leq \omega \leq \lambda_\tau + \mu_\tau$.

For marginal time slot $\chi$, there exist 2 situations: if $\omega = \lambda_\chi - \mu_\chi$ holds. To conclude, following equations hold:

$$a + b + c = T - 1, \tag{63a}$$
$$1 + 0 \times a + 2 \times b + 1 \times c = Ind, \tag{63b}$$
$$(-1) \times a + 1 \times b + 0 \times c + v_{i,\chi}^* = E_i. \tag{63c}$$

where (63a) implies the total connected time slots; (63b) indicates the number of elements before $\omega = O_{(F+1)}$ in the ordered set; (63c) is built to satisfy the energy constraint in constraint (60a). To combine (63a) - (63c), we can get:

$$b - a = F - T, \tag{63d}$$

$$F - T + v_{i,\chi}^* = E_i, \Rightarrow v_{i,\chi}^* = \frac{2e_i^r}{p_i^{\max}} - \left\lfloor \frac{2e_i^r}{p_i^{\max}} \right\rfloor \tag{63e}$$

form (63e), we prove $v_{i,\tau}^* \in [-1,0]$ to be true. Thus, $\omega = O_{(F+1)}$ is feasible to the KKT constraints.

Else, when $\omega = \lambda_\chi + \mu_\chi$, constraints (63a) and (63c) are still true, other equations change as:

$$1 + 0 \times a + 2 \times b + 1 \times c + 1 = Ind, \tag{63f}$$
$$b - a = Ind - T - 1, \tag{63g}$$

$$Ind - T - 1 + v_{i,\chi}^* = E_i, \Rightarrow v_{i,\chi}^* = \frac{2e_i^r}{p_i^{\max}} + 1 - \left\lceil \frac{2e_i^r}{p_i^{\max}} \right\rceil \tag{63h}$$

form (63h), we prove $v_{i,\tau}^* \in [0,1]$ to be true. Thus, $\omega = O_{(F+1)}$ is feasible to the KKT constraints.

To conclude, all EVs with the same $t_i^a$, $t_i^d$, $F_i^{VIG}$ values have the same $\omega = O_{(F+1)}$ solution to its KKT conditions (63). Briefly speaking, EVs will have the same charging preference (i.e., no-charging, half-maximal-charging, maximal-charging, marginal-charging) at any time slot.

Therefore, All EVs with the same $t_i^a$, $t_i^d$, $F_i^{VIG}$ values can be equivalent to a single EV while assuring the optimality. ∎

C. Proof of Lamma 2

The optimal scheduling of a bidirectional charging EV $j$, which can be obtained from the problem (12), denoted by $[(x_{j,\tau}^*, y_{j,\tau}^*, z_{j,\tau}^*), \forall \tau]$. Notice that $x_{j,\tau}^* * y_{j,\tau}^* = 0$ will always hold because of the positive battery degradation price $\psi$. Firstly, we make an equivalent transformation to name $w_{j,\tau} = x_{j,\tau} - y_{j,\tau}$; Therefore, $x_{j,\tau} = (|w_{j,\tau}| + w_{j,\tau})/2$, $y_{j,\tau} = (|w_{j,\tau}| - w_{j,\tau})/2$, Assume that parking horizon of EV $j$ is denoted as $\mathcal{T}_j := \{1, \cdots, T_j\}$. Then the optimization of EV $j$ in (12) can be rewritten as,

$$\min_{w_{j,\tau}, z_{j,\tau}} \sum_{\tau=1}^{T_j} \lambda_\tau w_{j,\tau} - \mu_\tau z_{j,\tau} + \frac{\psi}{2}\left(|w_{j,\tau}| - w_{j,\tau}\right)$$

Subject to: $\sum_{\tau=1}^{T_j} w_{j,\tau} = e_j^r$,

$$e_j^{low} \leq \sum_\tau w_{j,\tau} \leq e_j^{up} \quad \forall \tau \in \mathcal{T}_j$$
$$0 \leq z_{j,\tau} \leq p_j^{\max} \quad \forall \tau \in \mathcal{T}_j$$
$$|w_{j,\tau}| \leq p_j^{\max} - z_{j,\tau} \quad \forall \tau \in \mathcal{T}_j$$
(64)

Note that, equation $z_{j,\tau}^* = p_j^{\max} - |w_{j,\tau}^*|$ holds for any $\tau \in \mathcal{T}_j$ in (64). Because regulation market prices $\mu_\tau$ will always be larger than zero in the objective function. Therefore, we can eliminate $z_{j,\tau}$ by using $p_j^{\max} - |w_{j,\tau}|$ to replace it while assuring the optimality of the problem. Moreover, to simplify the parameters, we name $v_{j,\tau} = w_{j,\tau}/p_j^{\max}$ as well as ignore the constant term and constant factor in the objective function. Then, we have,

$$\min_{v_{j,\tau}} \mathcal{V} = \sum_{\tau=1}^{T_j} \left(\lambda_\tau - \frac{\psi}{2}\right) v_{j,\tau} + \left(\mu_\tau + \frac{\psi}{2}\right)|v_{j,\tau}|$$

Subject to: $\sum_{\tau=1}^{T_j} v_{j,\tau} = \frac{e_j^r}{p_j^{\max}}, \quad (\omega)$

$$\frac{e_j^{low}}{p_j^{\max}} \leq \sum_{t=1}^\tau v_{j,t} \quad (\alpha_\tau) \tag{65}$$

$$\sum_{t=1}^\tau v_{j,t} \leq \frac{e_j^{up}}{p_j^{\max}} \quad (\beta_\tau)$$

$$-1 \leq v_{j,\tau}, \quad (\rho_\tau)$$
$$v_{j,\tau} \leq 1. \quad (\theta_\tau)$$

One can write the Karush-Kuhn-Tucker conditions for the minimization problem in (65) as

$$\sum_{\tau=1}^{T_j} v_{j,\tau} = \frac{e_j^r}{p_j^{\max}}, \tag{66a}$$

$$\frac{e_j^{low}}{p_j^{\max}} \leq \sum_{t=1}^\tau v_{j,t} \quad \forall \tau \in \mathcal{T}_j \tag{66b}$$

$$\sum_{t=1}^\tau v_{j,t} \leq \frac{e_j^{up}}{p_j^{\max}} \quad \forall \tau \in \mathcal{T}_j \tag{66c}$$

$$-1 \leq v_{j,\tau} \quad \forall \tau \in \mathcal{T}_j \tag{66d}$$

$$v_{j,\tau} \leq 1 \quad \forall \tau \in \mathcal{T}_j \tag{66e}$$

$$\frac{\partial \mathcal{V}}{\partial v_{j,\tau}} - \omega + \left(\sum_{t=\tau}^{T_j} \beta_t - \alpha_t\right) + \theta_\tau - \rho_\tau = 0 \quad \forall \tau \in \mathcal{T}_j \tag{66f}$$

$$\omega\left(\sum_{\tau=1}^{T_j} v_{j,\tau} - \frac{e_j^r}{p_j^{\max}}\right) = 0, \tag{66g}$$

$$\alpha_\tau\left(\sum_{t=1}^\tau v_{j,t} - \frac{e_j^{low}}{p_j^{\max}}\right) = 0 \quad \forall \tau \in \mathcal{T}_j \tag{66h}$$

$$\beta_\tau\left(\sum_{t=1}^\tau v_{j,t} - \frac{e_j^{up}}{p_j^{\max}}\right) = 0 \quad \forall \tau \in \mathcal{T}_j \tag{66i}$$

$$\rho_\tau(v_{j,\tau}+1)=0 \quad \forall \tau \in \mathcal{T}_i \tag{66j}$$

$$\theta_\tau(v_{j,\tau}-1)=0 \quad \forall \tau \in \mathcal{T}_i \tag{66k}$$

$$\alpha_\tau \geq 0, \beta_\tau \geq 0, \rho_\tau \geq 0, \theta_\tau \geq 0 \quad \forall \tau \in \mathcal{T}_i \tag{66l}$$

Since the problem (65) only has linear constraints, and the subgradient of nonlinear objective function can be solved as,

$$\frac{\partial \mathcal{V}}{\partial v_{j,\tau}} = \begin{cases} \lambda_\tau - \mu_\tau - \psi & \text{when } v_{j,\tau} < 0 \\ [\lambda_\tau - \mu_\tau - \psi, \lambda_\tau + \mu_\tau] & \text{when } v_{j,\tau} = 0 \\ \lambda_\tau + \mu_\tau & \text{when } v_{j,\tau} > 0 \end{cases} \tag{67}$$

Note that equations $\{\rho_\tau *\theta_\tau =0: \forall \tau \in \mathcal{T}_j\}$ holds, because the two kinds of constraints will not be active simultaneously as shown in (66h), (66i); also, equations $\{\alpha_\tau *\beta_\tau=0: \forall \tau \in \mathcal{T}_j\}$ holds, because the two kinds of constraints will not be active simultaneously as shown in (66j) - (66k).

We would like to discuss situations where $\{\lambda_\tau<\mu_\tau+\psi: \forall \tau\}$ always hold. And, we need to prove $v^*_{j,\tau} \geq 0 \Leftrightarrow y^*_{j,\tau}=0$, Otherwise, if we assume there exists $v^*_{j,\gamma}<0$, then we have $\theta_\gamma=0, \beta_\gamma=0$. Combined (66f) and (67), we find:

$$\frac{\partial \mathcal{V}}{\partial v_{j,\gamma}}+\theta_\gamma-\rho_\gamma=(\lambda_\tau-\mu_\tau-\psi)-\rho_\gamma<0, \tag{68a}$$

$$\frac{\partial \mathcal{V}}{\partial v_{j,\gamma}}+\theta_\gamma-\rho_\gamma=\frac{\partial \mathcal{V}}{\partial v_{j,\gamma-1}}+\beta_{\gamma-1}-\alpha_{\gamma-1}+\theta_{\gamma-1}-\rho_{\gamma-1}, \tag{68b}$$

$$\frac{\partial \mathcal{V}}{\partial v_{j,\gamma}}+\theta_\gamma-\rho_\gamma=\frac{\partial \mathcal{V}}{\partial v_{j,\gamma+1}}-\beta_\gamma+\alpha_\gamma+\theta_{\gamma+1}-\rho_{\gamma+1}, \tag{68c}$$

In order to make the right side of equation (68b) smaller than 0, $v^*_{j,\gamma-1}$ must satisfy $v^*_{j,\gamma-1} \leq 0$; otherwise when $v^*_{j,\gamma-1} > 0$, $\frac{\partial \mathcal{V}}{\partial v_{j,\gamma-1}}=\lambda_{\gamma-1}+\mu_{\gamma-1}>0, \rho_{\gamma-1}=\alpha_{\gamma-1}=0$, and we can find the right side always larger than zero.

Similarly, to make the right side of equation (68c) smaller than 0, $v^*_{j,\gamma+1}$ must satisfy $v^*_{j,\gamma+1} \leq 0$; otherwise when $v^*_{j,\gamma+1} > 0$, $\frac{\partial \mathcal{V}}{\partial v_{j,\gamma+1}}=\lambda_{\gamma+1}+\mu_{\gamma+1}>0, \rho_{\gamma+1}=\beta_\gamma=0$, and we can find the right side always larger than zero.

In this manner, we will have $\{v^*_{j,1},\ldots,v^*_{j,\gamma-1}\} \leq 0$, and $v^*_{j,\gamma}<0$, and $\{v^*_{j,\gamma+1},\ldots,v^*_{j,T_j}\} \leq 0$; however, the constraint (66a) does not stand, since $e^r_j>0$. To conclude, $v^*_{j,\tau} \geq 0, \forall \tau$ is proved. Therefore, $p^{dis*}_{j,\tau}=0, \forall \tau$.

With the conclusion $v^*_{j,\tau} \geq 0, \forall \tau$, we can get $\alpha_\tau=\beta_\tau=0, \forall \tau$. Thus, equations (66f) can be rewritten as

$$\frac{\partial \mathcal{V}}{\partial v_{j,\tau}}-\omega+\theta_\tau-\rho_\tau=0 \quad \forall \tau \in \mathcal{T}_j \tag{69}$$

Similar to discussions in this Section A (62), we separate the analysis into 4 cases as follows,

1) for all time slots where $\lambda_\tau-\mu_\tau-\psi > \omega$, we have $\frac{\partial \mathcal{V}}{\partial v_{j,\tau}}-\omega=\rho_\tau-\theta_\tau>0$. Then, $\rho_\tau>0, \theta_\tau=0, v^*_{j,\tau}=-1$. From the above analysis, we know this case would not happen.

2) for all time slots where $\lambda_\tau+\mu_\tau<\omega$, we have $\frac{\partial \mathcal{V}}{\partial v_{j,\tau}}-\omega=$ $\rho_\tau-\theta_\tau<0$. Then, $\rho_\tau=0, \theta_\tau>0$. From (66k), we have:

$$v^*_{j,\tau}=1; \Rightarrow x^*_{j,\tau}=p^{max}_j, z^*_{j,\tau}=0. \tag{70}$$

3) for all time slots where $\lambda_\tau-\mu_\tau-\psi<\omega<\lambda_\tau+\mu_\tau$, we have $v^*_{j,\tau}=0; \Rightarrow x^*_{j,\tau}=0, z^*_{j,\tau}=p^{max}_j$. Otherwise, if $v^*_{j,\tau}>0$; $\frac{\partial \mathcal{V}}{\partial v_{j,\tau}}-\omega>0$. Then, we have $\rho_\tau>0, \theta_\tau=0, v^*_{j,\tau}=-1$, which is paradoxical; Else if $v^*_{j,\tau}<0$; $\frac{\partial \mathcal{V}}{\partial v_{i,\tau}}-\omega<0$. Then, we have $\rho_\tau=0, \theta_\tau>0$. $v^*_{j,\tau}=1$, which is also paradoxical.

4) for all time slots where $\lambda_\tau-\mu_\tau-\psi=\omega$, we have $v^*_{j,\tau} \in [-1,0]$, and it will not happen; for all time slots where $\lambda_\tau+\mu_\tau=\omega$, we have $v^*_{j,\tau} \in [0,1]$. If there exists two time slots (α,β) in which $\lambda_\alpha+\mu_\alpha=\omega=\lambda_\beta+\mu_\beta$, we can find $v^*_{j,\alpha}+v^*_{j,\beta}=v'^*_{j,\alpha}+v'^*_{j,\beta}$, and both of them are optimal solutions. Therefore, we could always leave at most one time slot $v^*_{j,\chi}$ to be non-integer, and χ is defined as the marginal slot of the j-th EV. ∎

*D. Proof of Theorem 2*

From the above discussion, we find that $\omega>\lambda_\tau-\mu_\tau-\psi$ with the price relationship $\{\lambda_\tau<\mu_\tau+\psi: \forall \tau\}$. Note that all EVs in the set (where $j=1, \cdots, M$) have the same $t^a_j$, $t^d_j$, $F^{V2G}_j$ values. For simplicity, we assume that the common parking horizon is denoted as $\mathcal{T}:=\{1,\cdots,T\}$ and the same charging flexibility index is denoted as *Ind*.

We sort T elements of set $\{\lambda_\tau+\mu_\tau\}, \forall \tau \in \mathcal{T}$ to get a new ascending ordered set $\{O_{(1)}, \ldots, O_{(T)}\}$. Next, we prove $\omega=O_{(F)}$ is the solution to the KKT conditions, where *Ind* is the same integer charging flexibility index.

From discussions shown in Section C, we can group *T-1* parking time slots (except marginal time slot χ) into 2 cases:

1) There are *a* time slots, where $v^*_{j,\tau}=1$, and $\lambda_\tau+\mu_\tau \leq \omega$;
2) There are *b* time slots, where $v^*_{j,\tau}=0$, and $\omega \leq \lambda_\tau+\mu_\tau$.

For marginal time slot χ, $\omega=\lambda_\chi+\mu_\chi$ holds. To conclude, following equations hold:

$$a+b=T-1, \tag{71a}$$

$$a=F-1, \tag{71b}$$

$$a+v^*_{j,\chi}=\frac{e^r_j}{p^{max}_j}. \Rightarrow v^*_{j,\chi}=\frac{e^r_j}{p^{max}_j}+1-\left\lceil \frac{e^r_j}{p^{max}_j} \right\rceil. \tag{71c}$$

where (71a) implies the total connected time slots; (71b) indicates the number of elements before $\omega=O_{(F)}$ in the ordered set; (71c) is built to satisfy the energy constraint in constraint (66a), we can find that $v^*_{j,\chi} \in [0,1]$. Thus, $\omega=O_{(F)}$ is feasible to the KKT constraints.

To conclude, all EVs with the same $t^a_j$, $t^d_j$, $F^{V2G}_j$ values, have the same $\omega=O_{(F)}$ solution to its KKT conditions. Briefly speaking, EVs will have the same charging preference (i.e., no-charging, maximal-charging, marginal-charging) at any time slot.

Therefore, All EVs with the same $t^a_j$, $t^d_j$, $F^{V2G}_j$ values can be equivalent to a single EV while assuring the optimality. ∎